\documentclass{aa}  

\usepackage{graphicx}
\usepackage{txfonts}

\def\ltsima{$\; \buildrel < \over \sim \;$}
\def\simlt{\lower.5ex\hbox{\ltsima}}
\def\gtsima{$\; \buildrel > \over \sim \;$}
\def\simgt{\lower.5ex\hbox{\gtsima}}

\def\cm2{\mbox{$\mbox{cm}^{-2}$}}
\def\cm3{\mbox{$\mbox{cm}^{-3}$}}
\def\h2{\mbox{$_{\mbox{\tiny H2}}$}}
\begin{document}

 \title{The role of molecular filaments in the origin of the prestellar core mass function 
and stellar initial mass function}


\titlerunning{Molecular filaments and the origin of the IMF}

   \author{Ph.~Andr\'e\inst{1}
          \and
          D.~Arzoumanian\inst{2, 3, 1}
          \and
          V.~K\"onyves\inst{4,1}
          \and
          Y.~Shimajiri\inst{5,1, 6}
          \and
          P.~Palmeirim\inst{3}                 
          }

   \institute{Laboratoire d'Astrophysique (AIM), CEA/DRF, CNRS, Universit\'e Paris-Saclay, Universit\'e Paris Diderot, Sorbonne Paris Cit\'e, 91191 Gif-sur-Yvette, France\\
              \email{pandre@cea.fr}\\
         \and
             Department of Physics, Graduate School of Science, Nagoya University, Furo-cho, Chikusa-ku, Nagoya 464-8602, Japan\\
                  \and
             Instituto de Astrof\'isica e Ci{\^e}ncias do Espa\c{c}o, Universidade do Porto, CAUP, Rua das Estrelas, PT4150-762 Porto, Portugal\\    
             \email{doris.arzoumanian@astro.up.pt}\\
         \and
         Jeremiah Horrocks Institute, University of Central Lancashire, Preston PR1 2HE, United Kingdom\\
         \and
             Department of Physics and Astronomy, Graduate School of Science and Engineering, Kagoshima University, 1-21-35 Korimoto, Kagoshima, Kagoshima 890-0065, Japan\\
         \and
             National Astronomical Observatory of Japan, Osawa 2-21-1, Mitaka, Tokyo 181-8588, Japan\\    
             }

   \date{Received May 17, 2019; accepted July 29, 2019}

 
  \abstract
{The origin of the stellar initial mass function (IMF) is one of the most debated issues in astrophysics.}
  %
{Here, we explore the possible link between the quasi-universal filamentary structure of star-forming molecular clouds 
and the origin of the IMF.}  
  %
{Based on our recent comprehensive study of filament properties from {\it Herschel} Gould Belt survey 
observations (Arzoumanian et al.), we derive, for the first time, a good estimate of the filament mass function (FMF) 
and filament line mass function (FLMF) in nearby molecular clouds. We use the observed FLMF to propose a simple
toy model for the origin of the prestellar core mass function (CMF), relying on gravitational fragmentation 
of thermally supercritical but virialized filaments.}
  %
{We find that the FMF and the FLMF have very similar shapes and are both consistent with a Salpeter-like 
power-law function (d$N$/dlog$M_{\rm line} \propto M_{\rm line}^{-1.5\pm0.1}$) in the regime 
of thermally supercritical filaments ($M_{\rm line} > 16\, M_\odot$/pc). 
This is a remarkable result since, in contrast, the mass distribution of molecular clouds and clumps is known to be 
significantly {\it shallower} than the Salpeter power-law IMF, with d$N$/dlog$M_{\rm cl} \propto M_{\rm cl}^{-0.7}$.
}    
  %
{Since the vast majority of prestellar cores appear to form in thermally transcritical or supercritical filaments, 
we suggest that the prestellar CMF and by extension the stellar IMF are at least partly inherited from the FLMF 
through gravitational fragmentation of individual filaments. 
}    


   \maketitle
%

\section{Introduction}
\label{sec:intro}

The origin of the stellar initial mass function (IMF) is a fundamental problem in modern astrophysics 
which remains highly debated \citep[e.g.][]{Offner+2014}. 
Two major features of the IMF are 1) a fairly robust power-law 
slope at the high-mass end \citep{Salpeter1955},  
and 2) a broad peak around $\sim 0.3\, M_\odot $ 
corresponding to a characteristic stellar mass scale 
\citep[e.g.][]{Larson1985}.
In recent years, the dominant 
theoretical model proposed to account for these features has been the ``gravo-turbulent 
fragmentation'' picture \citep[e.g.][]{Padoan+2002,Hennebelle+2008}, 
whereby the properties 
of supersonic interstellar turbulence lead to the Salpeter power law 
while gravity and thermal physics set the characteristic mass scale \citep[cf.][]{Larson2005}. 
This picture is deterministic in the sense that stellar masses are directly inherited from 
the distribution of prestellar core masses resulting from cloud fragmentation prior to protostellar collapse, 
in agreement with the observed similarity between 
the prestellar core mass function (CMF) and the system IMF \citep[e.g.][]{Motte+1998,Alves+2007,Konyves+2015}. 
In contrast, a major alternative view posits that stellar masses are essentially unrelated to initial prestellar core masses 
and result entirely from stochastic competitive accretion and dynamical interactions between 
protocluster seeds 
at the protostellar (Class~0/Class~I)  
stage of young stellar object evolution \citep[][]{Bonnell+2001,Bate+2003}.
Here, we discuss modifications to  
the gravo-turbulent picture based on  \emph{Herschel} results in nearby 
molecular clouds which emphasize the importance of filaments in the core/star formation process 
and potentially the CMF/IMF \citep[e.g.][]{Andre+2010}. 

\begin{figure*}
\centerline{\resizebox{0.48\hsize}{!}{\includegraphics[angle=0]{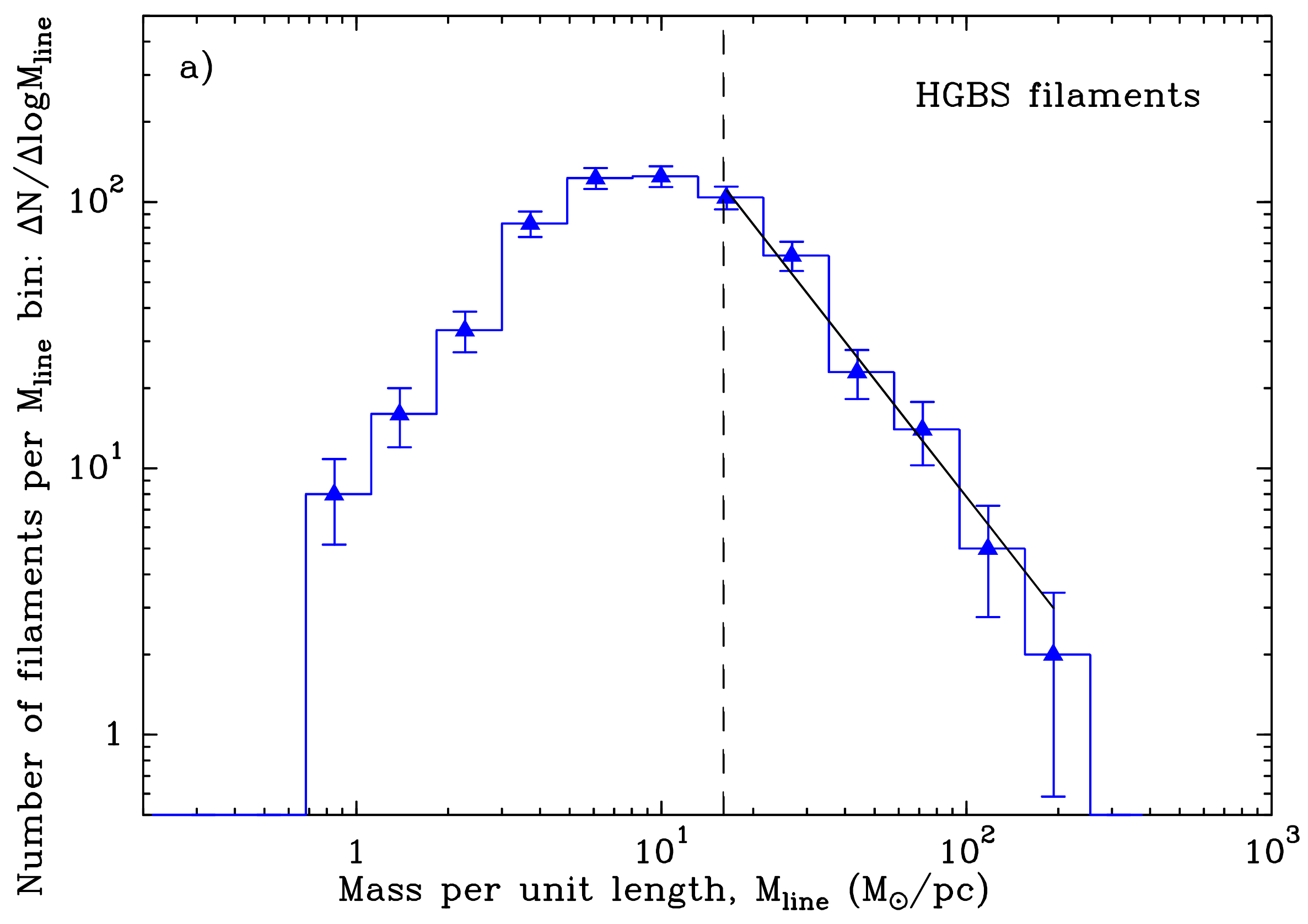}}
            \hspace*{0.3cm}
            \resizebox{0.48\hsize}{!}{\includegraphics[angle=0]{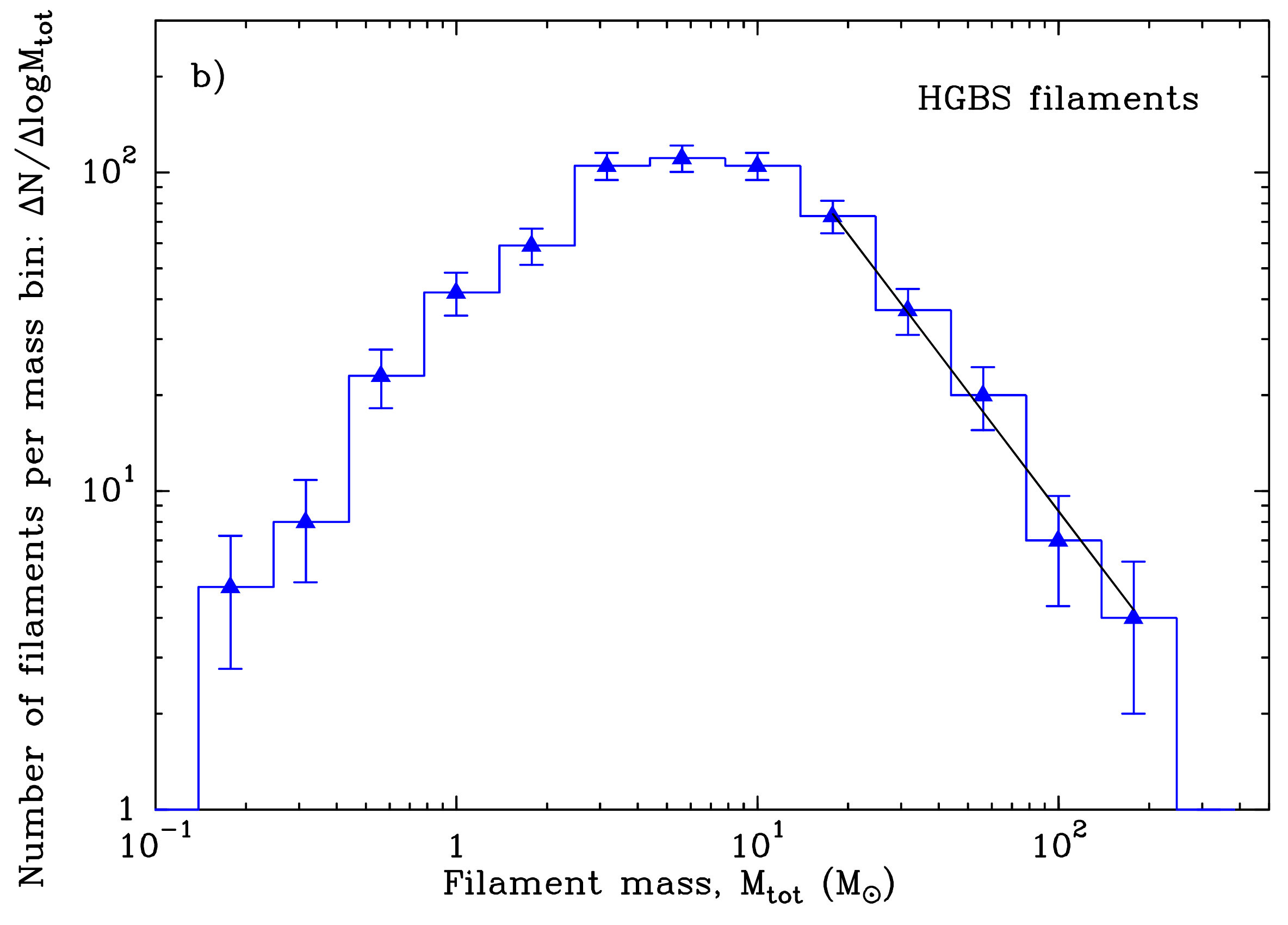}}}            
\caption{{\bf a)} Differential distribution of crest-averaged masses per unit length for the sample of 599 robust filaments identified by \citet{Arzoumanian+2018} 
in the {\it Herschel} GBS maps of eight nearby molecular clouds (IC5146, Orion~B, Aquila, Musca, Polaris, Pipe, Taurus L1495, 
and Ophiuchus). 
Above the critical mass per unit length $M_{\rm line, crit} \sim 16\, M_\odot$/pc (vertical dashed line), 
the filament sample is estimated to be $> 90\%$ complete (see text) and 
the distribution is well fitted by a Salpeter-like power law $\Delta N$/$\Delta$log$M_{\rm line} \propto M_{\rm line}^{-1.6\pm0.1}$ (solid line segment).
{\bf b)} Differential distribution of total masses for the same sample of filaments as in the left panel. 
At the high-mass end ($M_{\rm tot} > 15\, M_\odot$), the distribution of filament masses is well fitted by a Salpeter-like power law 
$\Delta N$/$\Delta$log$M_{\rm tot} \propto M_{\rm tot}^{-1.4\pm0.1}$  (solid line segment).
In both panels, the error bars correspond to $\sqrt{N}$ counting uncertainties. 
}
\label{fmf}
\end{figure*}

{\it Herschel} imaging observations have shown that filamentary structures are truly 
ubiquitous in the cold interstellar medium (ISM) of the Milky Way \citep[][]{Molinari+2010}, 
dominate the mass budget of Galactic molecular clouds at high ($\ga 10^4\, {\rm cm}^{-3} $) densities \citep[][]{Schisano+2014,Konyves+2015}, 
and feature a high degree of universality in their properties. 
In particular, detailed analysis of the radial column density profiles 
indicates that, at least in the nearby clouds of the Gould Belt, 
molecular filaments are characterized by a narrow distribution of crest-averaged inner widths with a typical full width at half maximum (FWHM) 
value $W_{\rm fil} \sim 0.1$~pc 
and a dispersion of less than a factor of $\sim \,$2 
\citep[][]{Arzoumanian+2011,Arzoumanian+2018, KochRosolowsky2015}.
Another major result from {\it Herschel} \citep[e.g.][]{Andre+2010,Konyves+2015,Marsh+2016} 
is that the vast majority ($>75\% $) of prestellar cores  
are found in dense,  ``transcritical'' or ``supercritical'' 
filaments for which the mass per unit length, $M_{\rm line}$, is close to or exceeds the critical line mass of nearly isothermal, long cylinders  
\citep[e.g.][]{Inutsuka+1997}, 
$M_{\rm line, crit} = 2\, c_{\rm s}^2/G \sim 16\, M_\odot$/pc,  
where $c_{\rm s} \sim 0.2$~km/s is the isothermal sound speed for 
molecular gas at $T \sim 10$~K. 
Moreover, most prestellar cores lie very close to the crests, i.e., within the inner 0.1~pc portion, of their parent filaments 
\citep[e.g.][]{Konyves+2018,Ladjelate+2019}. 
These findings support a filamentary paradigm in which 
low-mass star formation occurs in two main steps \citep[][]{Andre+2014,Inutsuka+2015}:
First, multiple large-scale compressions of cold interstellar material in supersonic MHD flows 
generates a cobweb of $\sim 0.1$-pc-wide filaments within sheet-like or shell-like molecular gas layers in the ISM; 
second, the densest molecular filaments fragment into 
prestellar cores (and then 
protostars) by gravitational instability near or above the critical line mass 
$M_{\rm line, crit} $, 
corresponding to $\Sigma_{\rm gas}^{\rm crit} \sim M_{\rm line, crit}/W_{\rm fil}  \sim 160\, M_\odot $/pc$^2$ 
in gas surface density ($A_V \sim  7.5$) or $n_{\rm H_2} \sim  2 \times 10^4\, {\rm cm}^{-3} $ in volume density. 
This paradigm differs from the classical gravo-turbulent picture \citep[][]{MacLowKlessen2004} 
in that it relies on the anisotropic formation of dense structures (such as shells, filaments, cores) in the cold ISM
and the unique properties of filamentary geometry \citep[cf.][]{Larson2005}. 

In the present paper, we exploit the results of our recent comprehensive study of filament properties from 
{\it Herschel} Gould Belt survey (HGBS) observations \citep{Arzoumanian+2018} and argue that 
the distribution of filament masses per unit length may directly connect to the CMF and by extension the IMF. 
Section~\ref{sec:obs} presents our observational results on the filament line mass function. 
Section~\ref{sec:discussion} discusses potential implications of these results for the origin of the prestellar CMF. 
Section~\ref{sec:concl} discusses the possible origin of the filament line mass function and 
concludes the paper.


\section{Observations of the filament line mass function}
\label{sec:obs}

\begin{figure*}
\centerline{\resizebox{0.45\hsize}{!}{\includegraphics[angle=0]{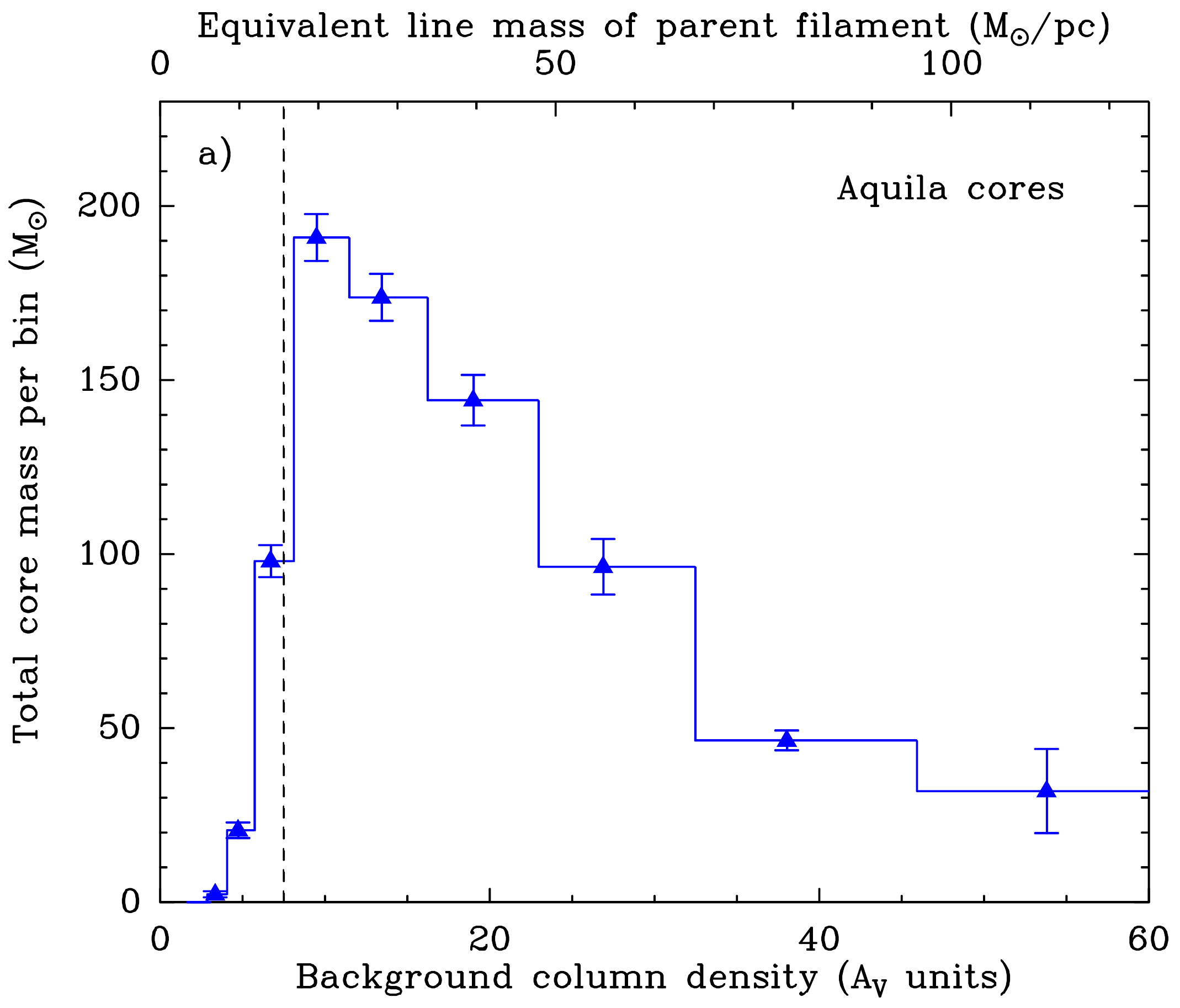}}
            \hspace*{0.5cm}
            \resizebox{0.45\hsize}{!}{\includegraphics[angle=0]{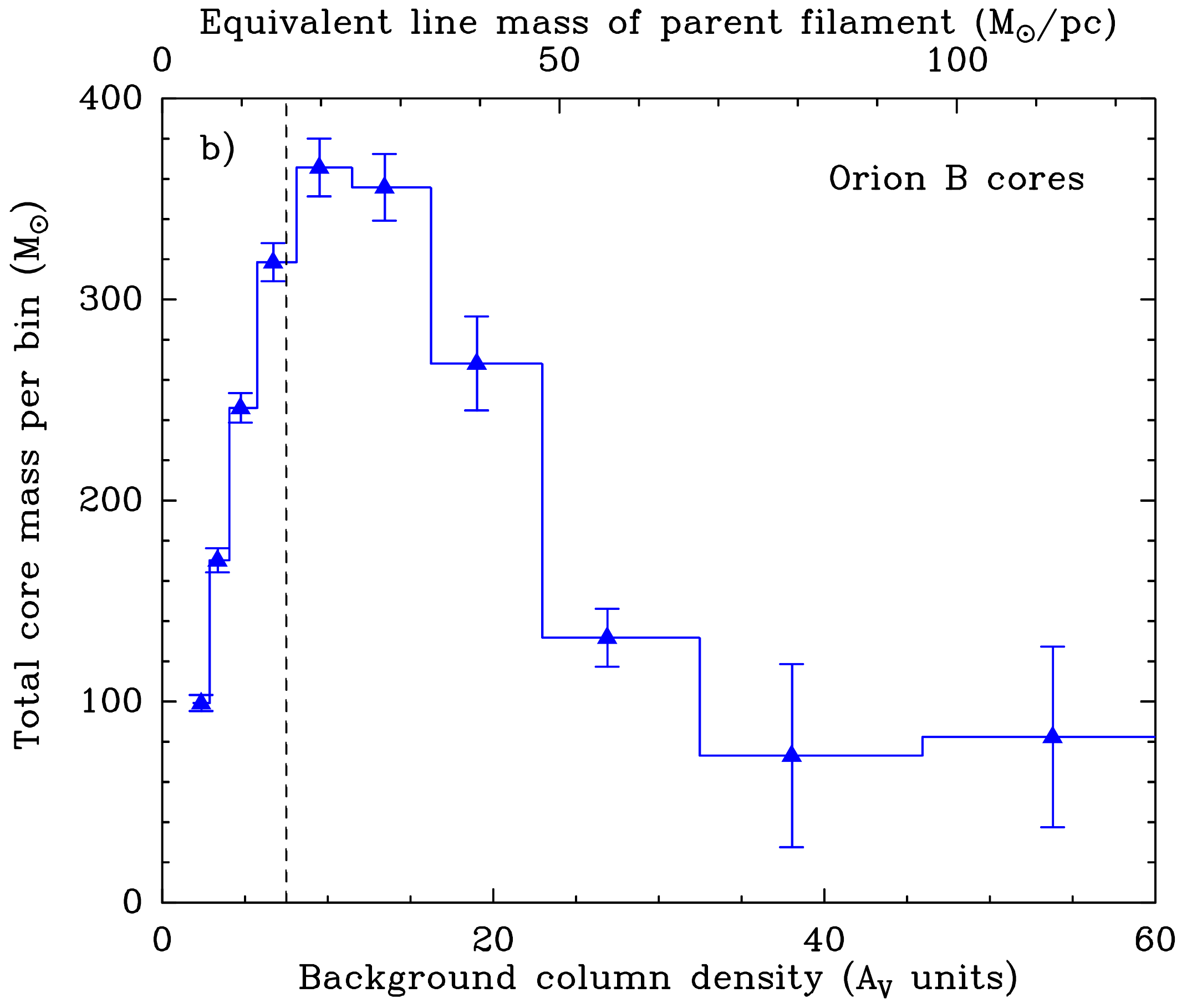}}}            
\caption{Total mass in the form of prestellar cores as a function of background column density (lower x-axis, in units of $10^{21}\, {\rm H_2\, cm^{-2}}$) 
or equivalent mass per unit length of the parent filament (upper x-axis, in units of $M_\odot $/pc) for the Aquila (left panel -- based on \citealp{Konyves+2015}) 
and Orion~B  (right panel -- adapted from \citealp{Konyves+2018}) clouds. 
The vertical dashed line marks the fiducial threshold for the formation of prestellar cores at a background $A_V = 7.5$, equivalent 
to a mass per unit length of $\sim 16\, M_\odot$/pc assuming parent filaments of 0.1~pc width.
}
\label{core_mass}
\end{figure*}

\citet{Arzoumanian+2018} recently presented a census of filament structures observed 
with {\it Herschel} in eight nearby regions covered by the HGBS: IC5146, Orion~B, Aquila, Musca, Polaris, Pipe, 
Taurus L1495, and Ophiuchus. 
Using the DisPerSE algorithm \citep{Sousbie2011} to trace filaments in the HGBS column density maps 
of these eight clouds\footnote{The corresponding column density maps and derived filament skeleton maps are 
available in fits format from:\\ http://gouldbelt-herschel.cea.fr/archives}, 
they identified a total of 1310 filamentary structures, including a selected sample of 599 robust filaments 
with aspect ratio (length/width) $> 3$ and central column density contrast 
$\delta \Sigma_{\rm fil}/\Sigma_{\rm cloud} > 30\% $ (where $\delta \Sigma_{\rm fil}$  is the background-subtracted gas surface density of the filament 
and $\Sigma_{\rm cloud}$ the surface density of the parent cloud). 
Performing an extensive set of tests on synthetic data, \citet[][see their Appendix A]{Arzoumanian+2018} estimated their selected sample 
of 599 filaments to be more than 95\% complete 
(and contaminated by less than 5\% of spurious detections) 
for filaments with column density contrast $\geq 100\% $. 
For reference, the column density contrast of isothermal model filaments in pressure equilibrium with 
their parent cloud is 
$<\delta \Sigma_{\rm fil}>/\Sigma_{\rm cloud}\approx 1.18\times \sqrt{f_{\rm cyl}/(1-f_{\rm cyl})}$, 
where 
$f_{\rm cyl} \equiv M_{\rm line}/M_{\rm line,crit} < 1 $ \citep[cf.][]{Fischera+2012}.\footnote{Equilibrium model filaments exist only for subcritical 
masses per unit length, i.e.,  $f_{\rm cyl} \leq 1$.}
Thermally transcritical filaments with $M_{\rm line,crit}/2  \la M_{\rm line} < M_{\rm line,crit} $ (i.e., $f_{\rm cyl} \ga 0.5 $) 
are therefore expected to have column density contrasts $\ga 100\%$, 
while thermally supercritical filaments with well-developed power-law density profiles reach column density contrasts $ >> 100\%$. 
The selected sample of \citet{Arzoumanian+2018} is thus estimated to be $> 95\%$ complete to thermally supercritical filaments 
with $M_{\rm line} > M_{\rm line,crit} \sim 16\, M_\odot$/pc. 

The differential distribution of average masses per unit length -- or {\it filament line mass function} (FLMF) --
derived   
from {\it Herschel} data for the 599 filaments of this sample is shown in Fig.~\ref{fmf}a. 
It can be seen that the FLMF 
is consistent with a power-law distribution in the supercritical mass per unit length regime (above $16\, M_\odot$/pc), 
$\Delta N$/$\Delta$log$M_{\rm line} \propto M_{\rm line}^{-1.59\pm0.07}$, 
at a Kolmogorov-Smirnov (K-S) significance level of 92\%. 
The error bar on the power-law exponent was derived by performing a non-parametric K-S test 
\citep[see, e.g.][]{Press+1992} on the cumulative distribution of masses per unit length N($>$$M_{\rm line}$), 
and corresponds to the range of exponents for which the K-S significance level is larger than 68\% 
(equivalent to $1\, \sigma $ in Gaussian statistics). 
Remarkably, the FLMF function observed above $M_{\rm line,crit} \sim 16\, M_\odot$/pc 
is very similar to the Salpeter power-law IMF \citep{Salpeter1955}, which scales as d$N$/dlog$M_\star$ $\propto$ $M_\star^{-1.35}$ in the same format. 

The right panel of Fig.~\ref{fmf} shows the distribution of {\it total} masses, integrated over filament length, for the same sample of filaments. 
As can be seen in Fig.~\ref{fmf}a, this {\it filament mass function} (FMF) is very similar in shape to the FLMF of Fig.~\ref{fmf}a, 
and is also consistent with Salpeter-like power-law distribution at the high-mass end ($M_{\rm tot} > 15\, M_\odot $),  
$\Delta N$/$\Delta$log$M_{\rm tot} \propto M_{\rm tot}^{-1.38 \pm0.10}$, 
at a Kolmogorov-Smirnov (K-S) significance level of 98\%. 
The similarity between the FMF and the FLMF is not surprising since  
$ M_{\rm tot} = M_{\rm line} \times L $ and the lengths $L$ of the filaments in the  \citet{Arzoumanian+2018} sample 
have an approximately lognormal distribution centered at about 0.5--0.6~pc (see Fig.~\ref{flf}a in Appendix~A), with no correlation with $M_{\rm line}$ 
(the linear Pearson correlation coefficient between  $L$ and $M_{\rm line}$ is $\lvert \rho \rvert < 8\% $). 
Accordingly, a strong linear correlation exists between $ M_{\rm tot} $ and  $ M_{\rm line} $ in the filament sample 
(correlation coefficient $ > 75\% $ -- see Fig.~\ref{flf}b in Appendix~A). 
We stress, however, that the estimated FMF shown in Fig.~\ref{fmf}b should be interpreted with caution and 
is not as robust as the FLMF of Fig.~\ref{fmf}a because filament-finding algorithms, such as DisPerSE used 
in the present analysis or \textsl{getfilaments} \citep{Menshchikov2013}, tend to break up filamentary structures into small filament segments.

\section{The role of filaments in the prestellar CMF}
\label{sec:discussion}

At least in terms of mass, most prestellar cores appear to form just above the fiducial column density ``threshold'' at $A_V \sim 7.5$, 
corresponding to marginally thermally supercritical filaments with $M_{\rm line} \ga 16\, M_\odot$/pc (\citealp{Konyves+2018} -- see also Fig.~\ref{core_mass}).
In the observationally-driven 
filamentary paradigm of star formation supported by {\it Herschel} 
results (see Sect.~\ref{sec:intro}), the dense cores making up 
the peak of the prestellar CMF -- presumably related to the peak of the IMF --
originate from gravitational fragmentation of filaments near the critical threshold for cylindrical gravitational instability \citep{Andre+2014}. 
In this picture, 
the characteristic prestellar core mass roughly 
corresponds to the local Jeans mass 
in transcritical or 
marginally supercritical filaments. 
The thermal Jeans or critical Bonnor-Ebert mass \citep[e.g.][]{Bonnor1956} is $M_{\rm BE, th} \approx 1.18\, c_{\rm s}^4 /(G^{3/2} P_{\rm cl}^{1/2}) $,  
where $P_{{\rm cl}}$ is the local pressure of the ambient cloud. The latter may be expressed as a function of cloud column density,  
$\Sigma_{\rm cl}$, as $P_{{\rm cl}} \approx 0.88~G~\Sigma_{\rm cl}^2$ \citep{McKeeTan2003}. 
Within a $\sim\, $0.1-pc-wide critical filament at $\sim \,$10\,K with $ M_{\rm line} \approx M_{\rm line, crit} \sim 16\, M_\odot \, {\rm pc}^{-1} $ 
and surface density $\Sigma_{\rm fil}  \approx \Sigma_{\rm gas}^{\rm crit} \sim 160\, M_\odot \, {\rm pc}^{-2} $ (see Sect.~\ref{sec:intro}), 
the local Bonnor-Ebert mass is thus: 

\begin{equation}
 M_{\rm BE, th}  \sim 1.3\, \frac{c_{\rm s}^4}{G^2 \Sigma_{\rm fil}} \sim 0.5\, M_\odot  \times \left(\frac{T}{10\, \rm K}\right)^2 \times  \left(\frac{ \Sigma_{\rm fil}}{160\, M_\odot \, {\rm pc}^{-2}}\right)^{-1}. 
\end{equation}

\noindent
This corresponds very well to the peak of the prestellar CMF at $ \sim 0.6\, M_\odot $ observed in the Aquila cloud \citep{Konyves+2015} 
and is also consistent within a factor $< 2$ with the CMF peak found with {\it Herschel} in other nearby regions such as Taurus L1495 \citep{Marsh+2016} 
or Ophiuchus \citep{Ladjelate+2019}.

The fragmentation of purely thermal, equilibrium filaments may be expected 
to result in a narrow (``$\delta$-like'') prestellar CMF sharply peaked 
at the median thermal Jeans mass \citep[see][]{Lee+2017}.
However, at least two effects 
contribute to broadening the observed CMF. 
First, the filament formation process through multiple large-scale compressions 
generates a field of initial density fluctuations within star-forming filaments 
\citep{Inutsuka2001,Inutsuka+2015}. 
Based on a study of the density fluctuations observed with {\it Herschel} along 
a sample of 80 subcritical or marginally supercritical filaments in three nearby clouds, 
\citet{Roy+2015} found that the power spectrum of line-mass fluctuations 
is well fitted by a power law, $P(k) \propto k^\alpha$ with $\alpha = {-1.6\pm0.3}$. 
This is consistent with the 1D power spectrum generated by subsonic Kolmogorov turbulence  ($\alpha = -5/3$). 
Starting from such an initial power spectrum,  
the theoretical analysis by \citet{Inutsuka2001} 
shows that the density perturbations quickly evolve 
-- in about two 
free-fall times or $\sim 0.5\,$Myr for a critical 0.1~pc-wide filament --
from a mass distribution similar to that of CO clumps \citep{Kramer+1998} 
to a population of protostellar cores whose mass distribution approaches the Salpeter power law 
at the high-mass end. 
This process alone is however unlikely to produce a CMF with a well-developed 
Salpeter-like power-law tail since very long filaments would be required.

\begin{figure}
\centerline{\resizebox{0.95\hsize}{!}{\includegraphics[angle=0]{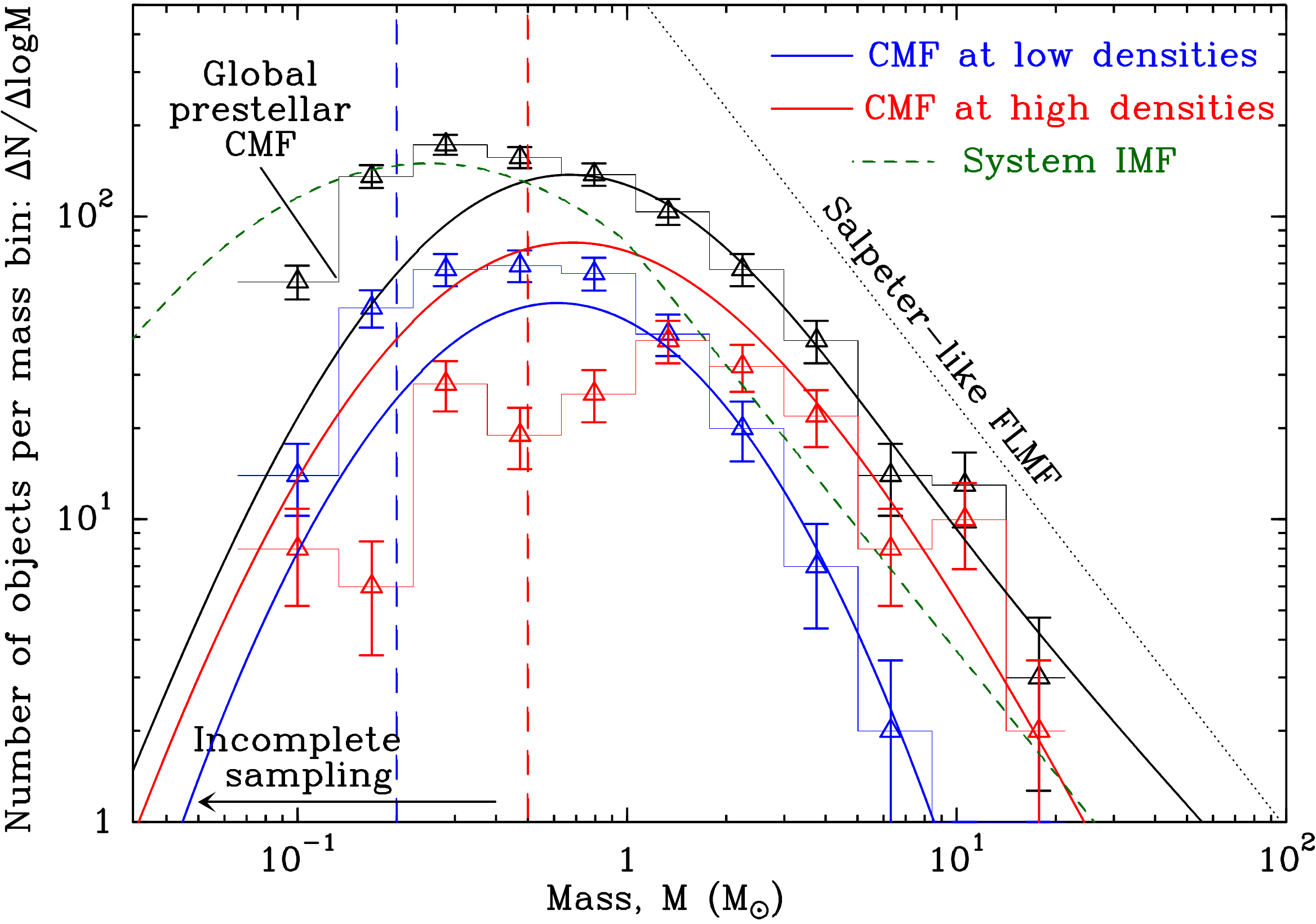}}}            
\caption{Comparison of the prestellar CMFs (in $\Delta N/\Delta$log$M$ format) expected in the toy model described in the text (solid curves) 
with the prestellar CMFs observed in the Orion~B cloud complex \citep{Konyves+2018} 
at low column densities  ($4 < A_V^{\rm back} < 7.5$, blue curve, data points, and histogram), 
higher column densities ($7.5 < A_V^{\rm back} < 21$, red curve, 
points, and histogram), 
and overall (all $A_V^{\rm back}$, black curve, 
points, and histogram).
The black dotted line displays the Salpeter-like power law FLMF, d$N$/dlog$M_{\rm line} \propto M_{\rm line}^{-1.4}$, 
assumed in the toy model and consistent with the observed FLMF in the supercritical regime (see Fig.~\ref{fmf}). 
The green dashed curve shows the system IMF from \citet[][]{Chabrier2005}.  
The two vertical dashed lines mark the estimated $80\% $ completeness limits of the {\it Herschel} census 
of prestellar cores in Orion~B at low and high background column densities respectively \citep[cf.][]{Konyves+2018}. 
The CMF extends to higher masses at higher 
column densities, i.e., higher $M_{\rm line}$ filaments 
in both the toy model and the observations.
}
\label{cmf_fil}
\end{figure}

A second broadening effect is due to the power-law distribution of filament masses per unit length (FLMF) in the supercritical regime (cf. Fig.~\ref{fmf}a). 
Given the typical filament width $W_{\rm fil} \sim 0.1\,$pc \citep{Arzoumanian+2011, Arzoumanian+2018} and the fact that 
thermally supercritical filaments are observed to be approximately 
virialized
with $ M_{\rm line} \sim  \Sigma_{\rm fil} \times W_{\rm fil}  \sim M_{\rm line, vir} \equiv 2\, c_{\rm s, eff}^2/G $,   
where $c_{\rm s, eff}$ is the one-dimensional velocity dispersion or  effective sound speed 
\citep{Fiege2000, Arzoumanian+2013}\footnote{Assuming rough equipartition between magnetic energy and kinetic energy,  
thermally supercritical filaments may also be close to magnetohydrostatic equilibrium, since the magnetic critical line mass $M^{\rm mag}_{\rm line, crit}$ 
may largely exceed $M_{\rm line, crit}$ \citep[cf.][]{Tomisaka2014}.}, 
the effective Bonnor-Ebert mass $ M_{\rm BE, eff}  \sim 1.3\, c_{\rm s, eff}^4 /(G^2 \Sigma_{\rm fil} $) scales roughly 
as  $\Sigma_{\rm fil}$ or $ M_{\rm line} $. 
At the same time, the thermal Bonnor-Ebert mass $ M_{\rm BE, th}$ scales roughly as  $\Sigma_{\rm fil}^{-1}$ or $ M_{\rm line}^{-1}$ (see Eq.~1). 
Hence, both higher- and lower-mass  cores may form in higher $ M_{\rm line} $ filaments.  
In agreement with this expected trend, dense cores of median mass $\sim 10\, M_\odot $,  
i.e., an order of magnitude higher that the peak of the prestellar CMF in low-mass nearby filaments (see above 
and Fig.~\ref{cmf_fil}), have recently been detected with ALMA in the 
NGC~6334 main filament which is an order of magnitude denser and more massive  ($ M_{\rm line}\,$$\sim \,$$1000\, M_\odot$/pc) 
than the Taurus B211/B213 filament and other Gould Belt filaments 
\citep{Shimajiri+2019b}. 
Furthermore, observations indicate that the prestellar CMF tends to be broader 
at higher ambient cloud column densities, i.e., in denser parent filaments (\citealp{Konyves+2018} -- see also Fig.~\ref{cmf_fil}). 
Since  the characteristic fragmentation mass $ M_{\rm BE, eff}$ scales linearly with $M_{\rm line} $, one may expect 
the Salpeter-like distribution of line masses 
observed above $M_{\rm line, crit}$ (cf. Fig.~\ref{fmf}a) 
to directly translate into a Salpeter-like power-law distribution of characteristic 
core masses. 
In 
detail, the global prestellar CMF 
results 
from the convolution of the CMF produced by individual filaments  
with the FLMF  \citep[cf.][]{Lee+2017}.

Based on the {\it Herschel} results and these 
qualitative considerations, we propose the following, observationally-driven 
quantitative scenario to illustrate the potential key role of the FLMF in the origin of the global prestellar CMF in molecular clouds. 
We assume that all prestellar cores form in thermally transcritical or supercritical (but virialized) filaments
and that the outcome of filament fragmentation depends only on the line mass of the parent filament. 
We denote by $ f_{M_{\rm line}}(m) \equiv {\rm d}N_{M_{\rm line}}/{\rm d\, log}\, m $ the differential CMF (per unit log mass, 
where $m$ represents core mass) in a filament of line mass $M_{\rm line}$. 
While the exact form of $ f_{M_{\rm line}}(m)$ is observationally quite uncertain, the foregoing 
arguments suggest that it should present a peak around the effective Bonnor-Ebert mass $ M_{\rm BE, eff}$ and may have 
a characteristic width scaling roughly as the ratio $ M_{\rm BE, eff}/M_{\rm BE, th} $.  
We thus make the minimal assumption that $ f_{M_{\rm line}}(m) $ follows a lognormal distribution centered 
at $ M_{\rm BE, eff}  (M_{\rm line}) $ and of standard deviation $ \sigma_{M_{\rm line}} (M_{\rm BE, eff}/M_{\rm BE, th}) $ in ${\rm log}\, m$:

\begin{equation}
 f_{M_{\rm line}}(m)  =  A  \times {\rm exp}\, \left(- \frac{({\rm log}\, m - {\rm log}\, M_{\rm BE, eff})^2}{2\,  \sigma_{M_{\rm line}}^2}\right). 
\end{equation} 

\noindent
We tested various simple functional forms for $ \sigma_{M_{\rm line}} (M_{\rm BE, eff}/M_{\rm BE, th}) $ 
and adopted $ \sigma_{M_{\rm line}}^2  = 0.4^2 +  0.3\, \left[{\rm log}\,(M_{\rm BE, eff}/M_{\rm BE, th})\right]^2 $  
as an illustrative fiducial form providing a reasonable good match to the observational constraints (see Fig.~\ref{cmf_fil} and Appendix~B). 

Denoting by $ g(M_{\rm line}) \equiv {\rm d}N/{\rm d\, log} \, M_{\rm line} $ the differential FLMF per unit log line mass, the global 
prestellar CMF per unit log mass  $\xi (m)  \equiv {\rm d}N_{\rm tot}/{\rm d\, log}\, m $ may be obtained as a weighted integration 
over line mass of the CMFs in individual filaments:

\begin{equation}
 \xi (m)  =   \int  f_{M_{\rm line}}(m)  \times w(M_{\rm line}) \times g(M_{\rm line}) \times d {\rm log} M_{\rm line}, 
\end{equation} 

\noindent
where $w(M_{\rm line}) \propto {\rm CFE}(M_{\rm line}) \times M_{\rm line} \times L $ represents 
the relative weight as a function of $M_{\rm line}$,  
${\rm CFE}(M_{\rm line}) $ is the prestellar core formation efficiency,  and $L$ the filament length. 
The results of Sect.~\ref{sec:obs} suggest that the FLMF is a power law 
$ g(M_{\rm line}) \propto M_{\rm line}^{-\alpha} $ with $\alpha \approx 1.4 $. 
As $L$ and $M_{\rm line}$ are not correlated in the filament sample of \citet{Arzoumanian+2018} (cf. Sect.~\ref{sec:obs}), 
we here adopt $L = {\rm constant} \sim 0.55\,$pc for simplicity (see Fig.~\ref{flf} in Appendix~A). 
Observationally, ${\rm CFE}(M_{\rm line}) $ exhibits a sharp transition 
between a regime of negligible prestellar core formation efficiency at $M_{\rm line} << M_{\rm line,crit} $ and 
a regime of roughly constant core formation efficiency $\sim \,$~15--20\% at $M_{\rm line} >>M_{\rm line,crit} $ (see Sect.~\ref{sec:intro}). 
Following \citet{Konyves+2015}, we describe this transition as a smooth step function of the form 
${\rm CFE}(M_{\rm line})  = {\rm CFE}_{\rm max} \times [1\, -\, {\rm exp}\, (1 - 2\,M_{\rm line}/M_{\rm line, crit}) ] $
with ${\rm CFE}_{\rm max} = 15\% $. 

The global prestellar CMF expected in the framework of this toy model, as well as the CMFs expected 
in thermally transcritical filaments and slightly supercritical filaments, are shown in Fig.~\ref{cmf_fil} as a black solid,  
blue solid, and 
red solid curve, respectively. For comparison, the black, blue, 
and red histograms with error bars 
represent the corresponding CMFs observed with {\it Herschel} 
in Orion~B \citep{Konyves+2018}. A good, overall agreement can be seen. 
Most importantly, it can 
be seen in Fig.~\ref{cmf_fil} that the global prestellar CMF approaches the power-law shape 
of the FLMF at the high-mass end.
We stress that the empirical toy model described here 
is only meant to quantify the 
links between the FLMF and the CMF/IMF.
It may also provide useful guidelines to help develop a self-consistent physical model 
for the origin of the CMF/IMF in filaments in the future. 

\section{Concluding remarks}
\label{sec:concl}

Our discussion of the {\it Herschel} observations in Sect.~\ref{sec:obs} indicates that both the filament line mass function (FLMF)
and the filament mass function (FMF) are consistent with a steep, Salpeter-like power-law (d$N$/dlog$M_{\rm line} \propto M_{\rm line}^{-1.6}$ 
and d$N$/dlog$M_{\rm tot} \propto M_{\rm tot}^{-1.4}$, respectively) 
in the regime of thermally supercritical filaments ($M_{\rm line} > 16\, M_\odot$/pc). 
This is a remarkable result since, in contrast, the mass distribution of molecular clouds and clumps is observed to be 
significantly {\it shallower} than the Salpeter power-law IMF, namely d$N$/dlog$M_{\rm cl} \propto M_{\rm cl}^{-0.7}$ \citep[][]{Blitz1993,Kramer+1998}. 
Theoretically, the latter is reasonably well understood in terms of the mass function of both ``bound objects
on the largest self-gravitating scale'' \citep{Hopkins2012a} and non-self-gravitating structures \citep{Hennebelle+2008}  
generated by supersonic interstellar turbulence. 
Thus, filamentary structures in molecular clouds appear to differ from standard clumps in a fundamental way
and may represent the key evolutionary step at which the steep slope of the prestellar CMF (and by extension that of the stellar IMF) 
originates (see Sect~\ref{sec:discussion}).

In the context of the filament paradigm summarized in Sect.~\ref{sec:intro},
we speculate that the observed FLMF arises from a combination of two effects. 
First, a spectrum of large-scale compression flows in the cold ISM produces 
a network of filamentary structures with an initial line mass distribution d$N$/dlog$M_{\rm line} \propto M_{\rm line}^{-1}$, 
determined by the power spectrum of interstellar turbulence (K. Iwasaki, private communication). 
Indeed, turbulence is known to generate essentially self-similar, fractal structure in interstellar clouds 
\citep[e.g.][]{Larson1992,Elmegreen+1996}, and this leads to a mass distribution of sub-structures with 
equal mass contribution per logarithmic interval of mass, i.e., d$N$/dlog$M \propto M^{-1}$, independent 
of the fractal dimension \citep[][]{Elmegreen1997,Padoan+2002}. 
Second, thermally supercritical filaments accrete mass from their parent molecular cloud \citep[][]{Arzoumanian+2013,Shimajiri+2019} 
due to their gravitational potential $\propto G\, M_{\rm line} $ \citep[][]{HennebelleAndre2013}. 
Therefore, they grow in mass per unit length at a rate $\dot{M}_{\rm line} \propto \sqrt{G\, M_{\rm line}} $ 
on a characteristic timescale $\tau_{\rm acc} = M_{\rm line}/\dot{M}_{\rm line} \propto \sqrt{M_{\rm line}} $, 
while fragmenting and forming cores on a comparable timescale \citep[cf.][]{Heitsch2013}. 
The accretion timescale is 
on the order of 1--2~Myr for a Taurus-like filament with $M_{\rm line} \sim 50\, M_\odot $/pc \citep[][]{Palmeirim+2013}.
As shown in Appendix~C, 
starting from an initial line mass spectrum d$N$/dlog$M_{\rm line} \propto M_{\rm line}^{-1}$, 
this accretion process leads to a steepening of the distribution of supercritical masses per unit length 
on a similar timescale 
(Fig.~\ref{model_flmf_nodecay}), 
and thus to a reasonable agreement with the observed FLMF (see Fig.~\ref{model_flmf}b).

Given the empirical toy model of Sect.~\ref{sec:discussion} for the CMF produced by a collection of molecular filaments 
and its reasonably good match to observations (Fig.~\ref{cmf_fil}), we conclude that the filament paradigm for star formation
provides  a promising conceptual framework for understanding the origin of the prestellar CMF and by extension the stellar IMF.

\begin{acknowledgements}
We are grateful to S. Inutsuka and P. Hennebelle for stimulating discussions. 
This work has received support from the European Research Council under the European Union's Seventh Framework Programme 
(ERC Advanced Grant Agreement no. 291294 - ORISTARS). 
We also acknowledge support from the French national programs of CNRS/INSU on stellar and ISM physics (PNPS and PCMI). 
DA and PP acknowledge support from FCT/MCTES through Portuguese national funds (PIDDAC) by  grant UID/FIS/04434/2019. 
PP also acknowledges support from fellowship SFRH/BPD/110176/2015 funded by FCT (Portugal) and POPH/FSE (EC).
YS is supported by NAOJ ALMA Scientific Research Grant Numbers 2017-04A. 
The present study has made use of data from the Herschel Gould Belt survey (HGBS) project (http://gouldbelt-herschel.cea.fr). 
The HGBS is a Herschel Key Programme jointly carried out by SPIRE Specialist Astronomy Group 3 (SAG 3), scientists 
of several institutes in the PACS Consortium (CEA Saclay, INAF-IFSI Rome and INAF-Arcetri, KU Leuven, MPIA Heidelberg), 
and scientists of the Herschel Science Center (HSC). 
\end{acknowledgements}

%
%

\bibliographystyle{aa}
\bibliography{filaments_ref}


\begin{appendix}

\section{Distribution of filament lengths}\label{App1}

\begin{figure}[!htp]
\centerline{\resizebox{0.95\hsize}{!}{\includegraphics[angle=0]{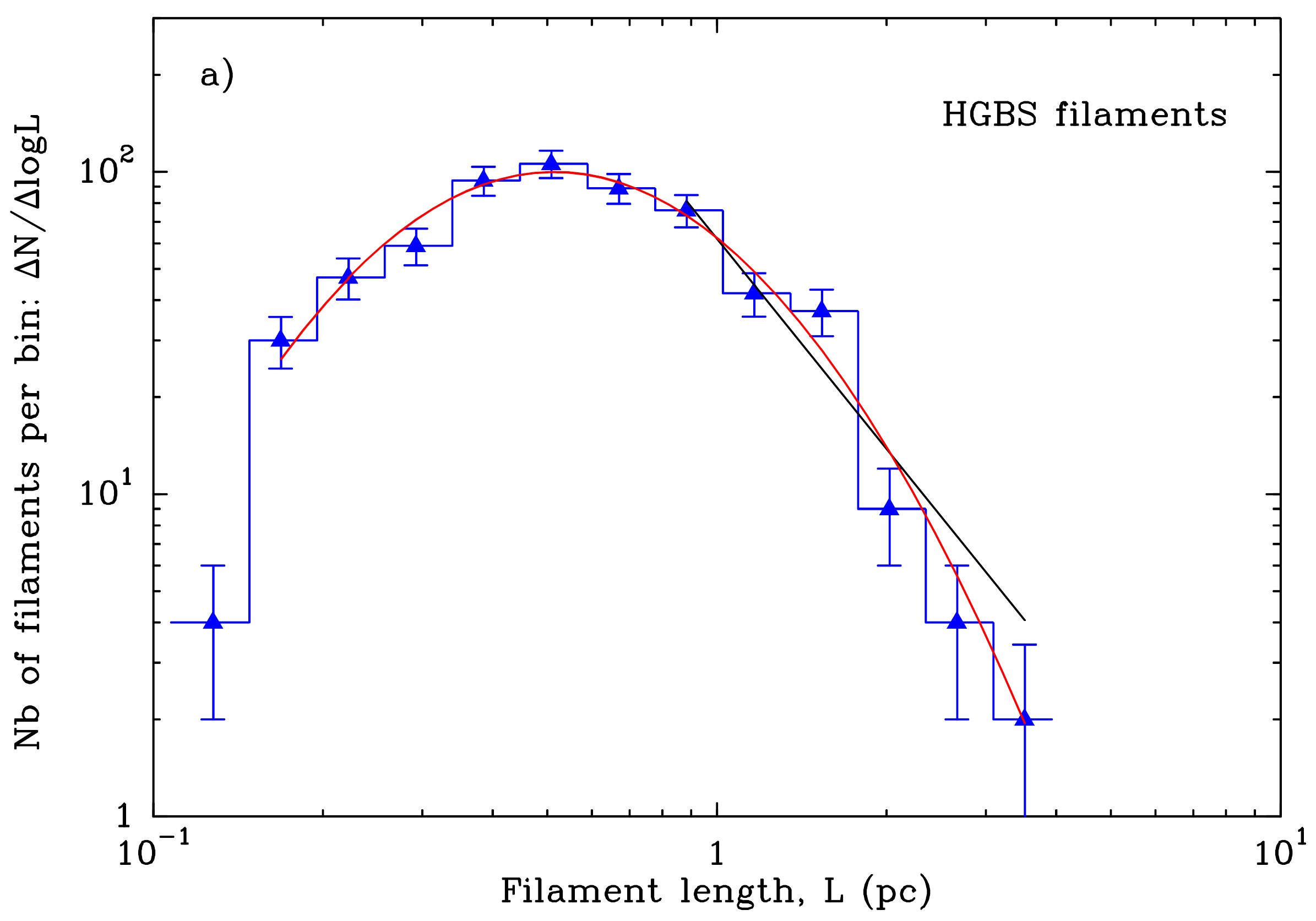}}}
            \vspace*{0.3cm}
\centerline{\resizebox{0.95\hsize}{!}{\includegraphics[angle=0]{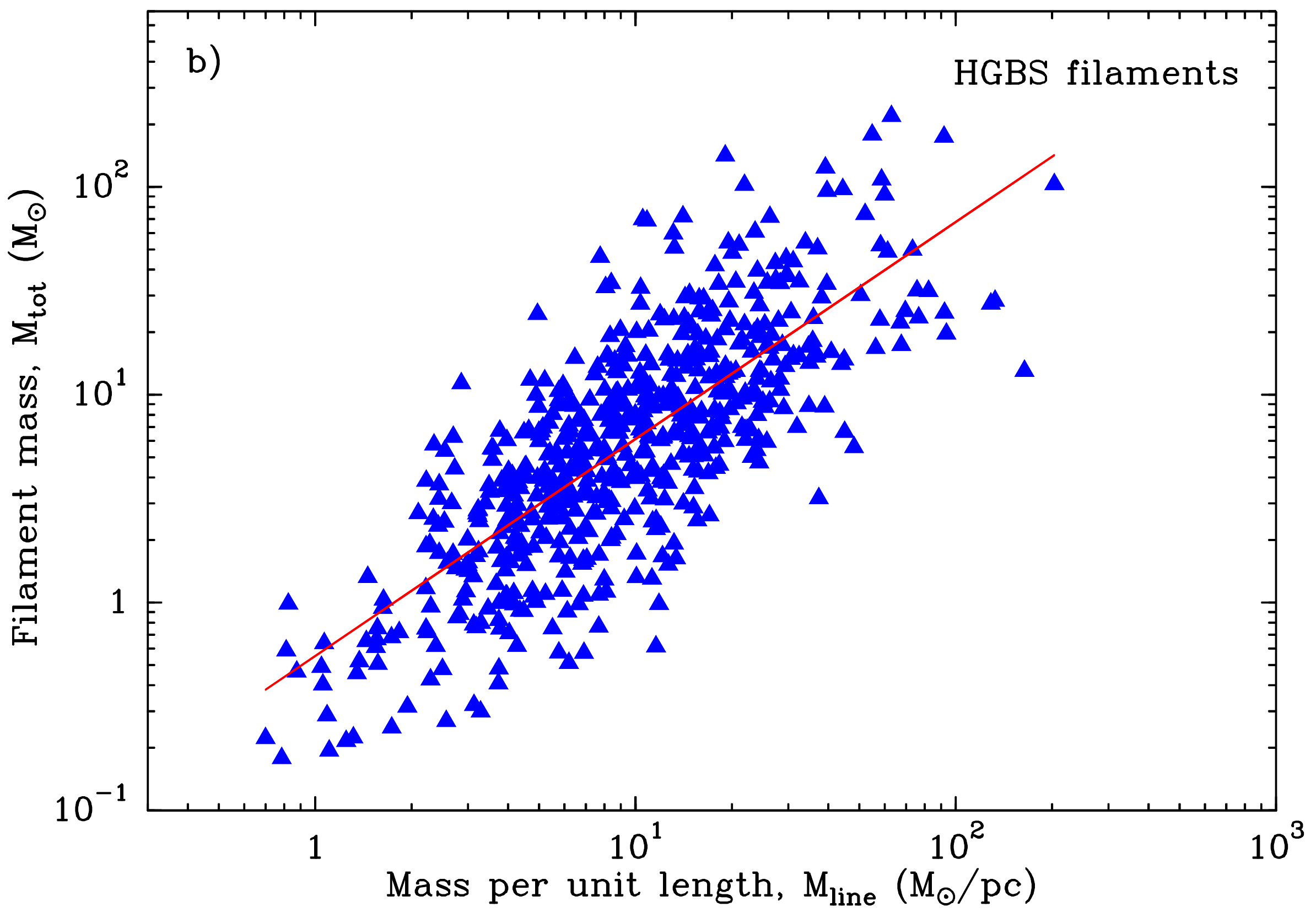}}}            
\caption{{\bf a)} Differential distribution of lengths for the sample of 599 robust HGBS filaments identified by \citet{Arzoumanian+2018}. 
The red curve shows a lognormal fit to the overall distribution, the black solid line segment a power-law fit for filaments longer than 0.9~pc. 
{\bf b)}~Plot of total mass $M_{\rm tot} $ against average mass per unit length $M_{\rm line}$ for the same sample of filaments as in the top panel. 
A strong linear correlation is observed between log\,$M_{\rm tot} $ and log\,$M_{\rm line}$ (with a Pearson correlation coefficient of $\sim \,$77\%).
The red line shows the best-fit linear relation, $M_{\rm tot} = M_{\rm line} \times L_{\rm eff} $, consistent with a typical effective length 
$L_{\rm eff} \sim 0.55\, $pc 
in the filament sample of \citet{Arzoumanian+2018}.
}
\label{flf}
\end{figure}

In this Appendix, we show the distribution of filament lengths in the \citet{Arzoumanian+2018} 
sample (Fig.~\ref{flf}a)  and the linear correlation between filament mass 
and filament mass per unit length (Fig.~\ref{flf}b), consistent with a roughly uniform 
length $L_{\rm eff} \sim 0.55\, $pc independent of $M_{\rm line}$.

\section{Observational constraints on the core mass function in individual filaments }\label{App2}

\begin{figure}[!htp]
\centerline{\resizebox{0.95\hsize}{!}{\includegraphics[angle=0]{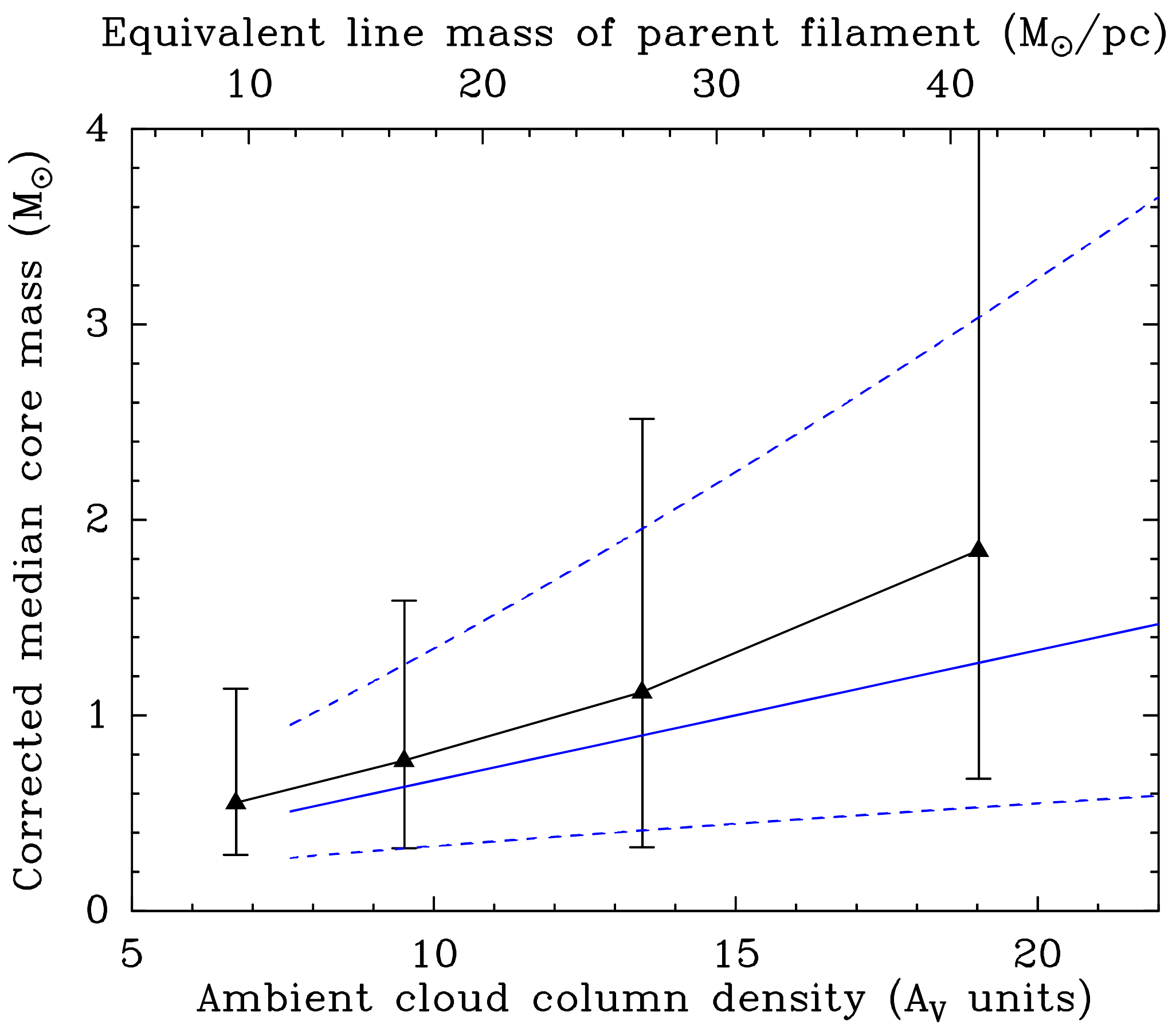}}} 
\caption{Median prestellar core mass versus background column density as observed in the Orion~B region after correction 
for incompleteness effects (black triangles, from \citealp{Konyves+2018}), 
compared to the prediction of the toy model described in Sect.~\ref{sec:discussion}  (blue solid line).
The error bars correspond to the inter-quartile range in observed masses for each bin of background column density. 
The two dashed blue lines mark the inter-quartile range expected in the context of the toy model (see Eq.~2).
}
\label{med_mcore_av}
\end{figure}

The form of the prestellar CMF produced by a single filament of line mass $M_{\rm line}$, 
denoted $ f_{M_{\rm line}}(m)$ in the text, is the most uncertain element of the empirical 
toy model described in Sect.~\ref{sec:discussion} for the CMF/IMF. 
For statistical reasons, observational estimates of CMFs in individual filaments are difficult owing 
to the relatively low number of cores per filament (see, however, the promising ALMA results of \citealp{Shimajiri+2019b} 
for the massive filament in NGC~6334).
Observations nevertheless indicate that the median prestellar core mass increases roughly linearly 
with the line mass of the parent filament and that the dispersion in core masses also increases 
with $M_{\rm line}$ \citep[][see also Fig.~\ref{med_mcore_av}]{Konyves+2018}.
In agreement with this observational trend, the qualitative arguments presented in Sect.~\ref{sec:discussion} 
suggest that the characteristic prestellar core mass should scale with the effective Bonnor-Ebert mass $ M_{\rm BE, eff}$ 
in the parent filament and that the dispersion in core masses may scale with the ratio $ M_{\rm BE, eff}/M_{\rm BE, th} $. 
The blue lines in Fig.~\ref{med_mcore_av} show how the median core mass and the dispersion in core masses 
vary with $M_{\rm line}$ in the toy model of Sect.~\ref{sec:discussion}, which assumes a lognormal shape  
for $ f_{M_{\rm line}}(m)$ with standard deviation $ \sigma_{M_{\rm line}}  = \sqrt{0.4^2 +  0.3\, \left[{\rm log}\,(M_{\rm BE, eff}/M_{\rm BE, th})\right]^2} $. 
The latter expression for  $ \sigma_{M_{\rm line}}$ corresponds to the quadratic sum of two terms: 
the first term represents the intrinsic spread in the core masses generated by transcritical filaments (which have $M_{\rm BE, eff}/M_{\rm BE, th} \sim 1 $), 
while the second term represents the spread in characteristic fragmentation masses within supercritical 
but virialized filaments (which have $M_{\rm BE, eff}/M_{\rm BE, th} > 1 $ -- see Sect.~\ref{sec:discussion}). 
It can be seen in Fig.~\ref{med_mcore_av} that these simple assumptions about $ f_{M_{\rm line}}(m)$  and $ \sigma_{M_{\rm line}} $ 
match the observational constraints reasonably well. 

\begin{figure}[!htbp]
\centerline{\resizebox{0.95\hsize}{!}{\includegraphics[angle=0]{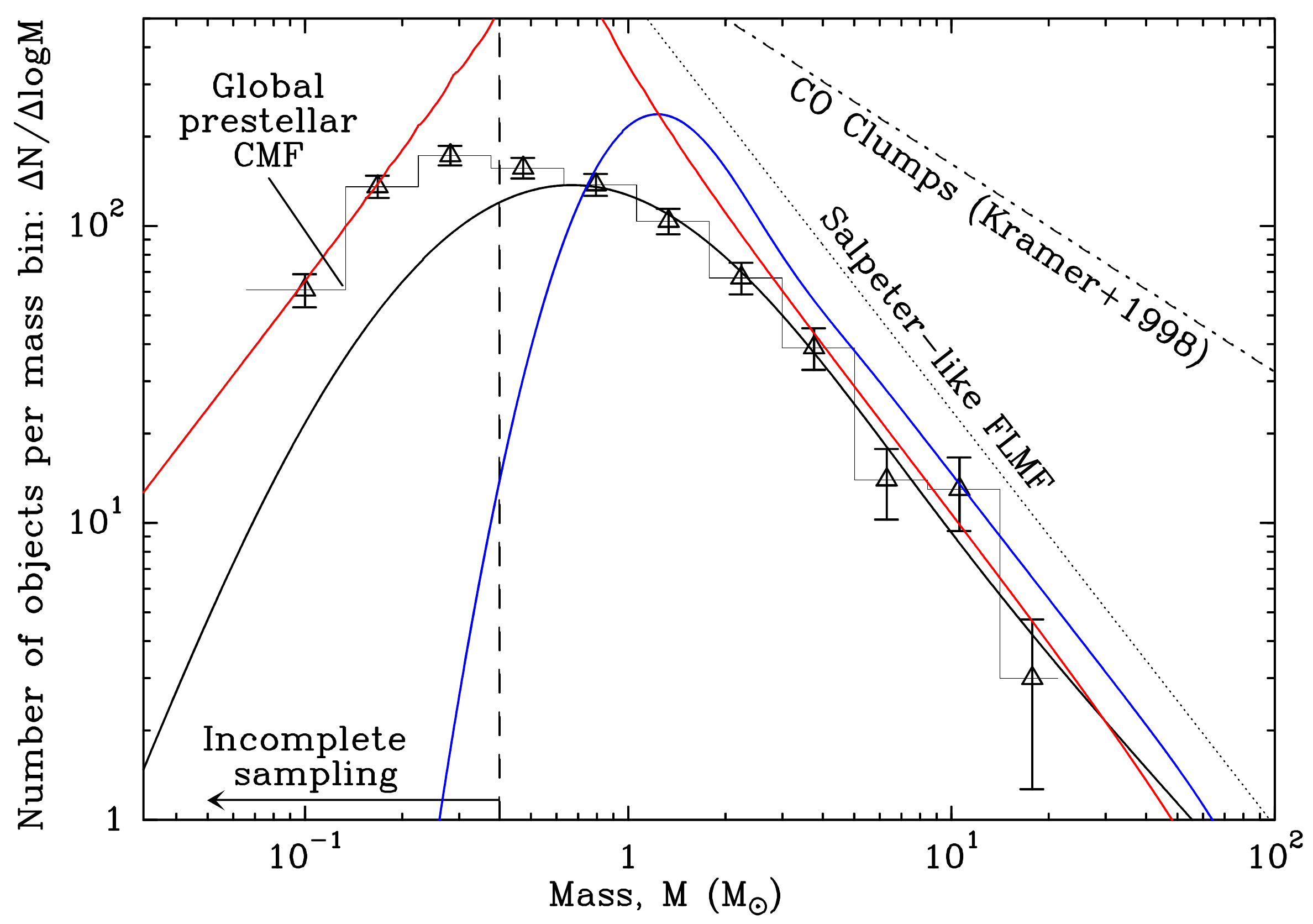}}}            
\caption{Comparison of the prestellar CMF observed in Orion~B (black triangular data points 
and histogram from \citealp{Konyves+2018} -- see also Fig.~\ref{cmf_fil}) 
with the global prestellar CMFs 
expected in the toy model of Sect.~\ref{sec:discussion}, under three assumptions 
about the shape of the CMF generated by a single filament  
of line mass $M_{\rm line}$: i) lognormal $ f_{M_{\rm line}}(m)$ distribution (Eq.~2) with $ \sigma_{M_{\rm line}}  = \sqrt{0.4^2 +  0.3\, \left[{\rm log}\,(M_{\rm BE, eff}/M_{\rm BE, th})\right]^2} $ 
(fiducial case -- black curve) , ii) lognormal  $ f_{M_{\rm line}}(m)$ distribution (Eq.~2) with fixed $ \sigma_{M_{\rm line}}  = 0.2 $ independent of $M_{\rm line}$ 
(blue curve), and iii) broken power law, $ f_{M_{\rm line}}(m) \propto m^6$ for $m <  M_{\rm BE, th}$, $ f_{M_{\rm line}}(m) = {\rm constant} $ 
for $M_{\rm BE, th} \le m <  M_{\rm BE, eff}$, $ f_{M_{\rm line}}(m) \propto m^{-4}$ for $m >  M_{\rm BE, eff}$ (red curve).
For reference, the black dotted line displays the Salpeter-like power law FLMF, d$N$/dlog$M_{\rm line} \propto M_{\rm line}^{-1.4}$, 
observed in the supercritical line mass regime (see Fig.~\ref{fmf}), and the black dash-dotted line shows the typical mass distribution 
of CO clumps \citep{Kramer+1998}. 
}
\label{cmf_fil_models}
\end{figure}

We also stress that the high-mass end of the global prestellar CMF in our toy model is primarily driven by the power-law shape 
of the FLMF and depends only weakly on the detailed form assumed for $ f_{M_{\rm line}}(m) \equiv {\rm d}N_{M_{\rm line}}/{\rm d\, log}\, m $.
This is illustrated in Fig.~\ref{cmf_fil_models} which shows the model global CMFs for three different assumptions 
about $ f_{M_{\rm line}}(m)$, 
compared to the prestellar CMF observed in Orion~B \citep[][see also Fig.~\ref{cmf_fil}]{Konyves+2018}. 
It can be seen that the three model CMFs are consistent with a  Salpeter-like power law at the high-mass end 
and only differ significantly at the low-mass end.

\newpage

\section{A toy accretion model for the filament line mass function}\label{App3}

As mentioned in Sect.~\ref{sec:concl}, thermally supercritical filaments are believed to accrete mass from their parent molecular cloud \citep[][]{Arzoumanian+2013,Shimajiri+2019} 
owing to their gravitational potential $\propto G\, M_{\rm line} $ \citep[][]{Heitsch2013,HennebelleAndre2013}. 
This leads to an accretion rate $\dot{M}_{\rm line} \propto \sqrt{G\, M_{\rm line}} $ \citep[cf.][]{Palmeirim+2013} and therefore to a simple differential 
equation of the form: 

\begin{equation}
 \frac{{\rm d} M_{\rm line} }{{\rm d} t}  =   A\, M_{\rm line}^{1/2} , 
\end{equation} 

\noindent
where $A$ is a positive constant. 
This equation can be easily integrated to give the time evolution of the line mass due to gravitational accretion:

\begin{equation}
M_{\rm line}(t) = \left[M_{\rm line}(0)^{1/2} + \frac{A}{2}\, t\,  \right]^{\,2}.
\end{equation} 

\noindent
If we choose to express time $\tilde{t} $ in units of the time needed 
to increase the line mass of an initially critical filament 
by a factor of 4, Eq.~(C.2) can be written in the form:

\begin{equation}
M_{\rm line}(\tilde{t}) = \left[M_{\rm line}(0)^{1/2} + M_{\rm line, crit}^{1/2}\, \tilde{t}\,  \right]^{\,2}.
\end{equation} 

\noindent 
In these units, the characteristic instantaneous accretion timescale is: 

\begin{equation}
\tilde{\tau}_{\rm acc} = M_{\rm line}/\dot{M}_{\rm line}  =  \frac{1}{2}\, \left(\frac{M_{\rm line}}{M_{\rm line, crit}} \right)^{1/2} \propto \sqrt{M_{\rm line}} . 
\end{equation}

\begin{figure}[!htp]
\centerline{\resizebox{0.95\hsize}{!}{\includegraphics[angle=0]{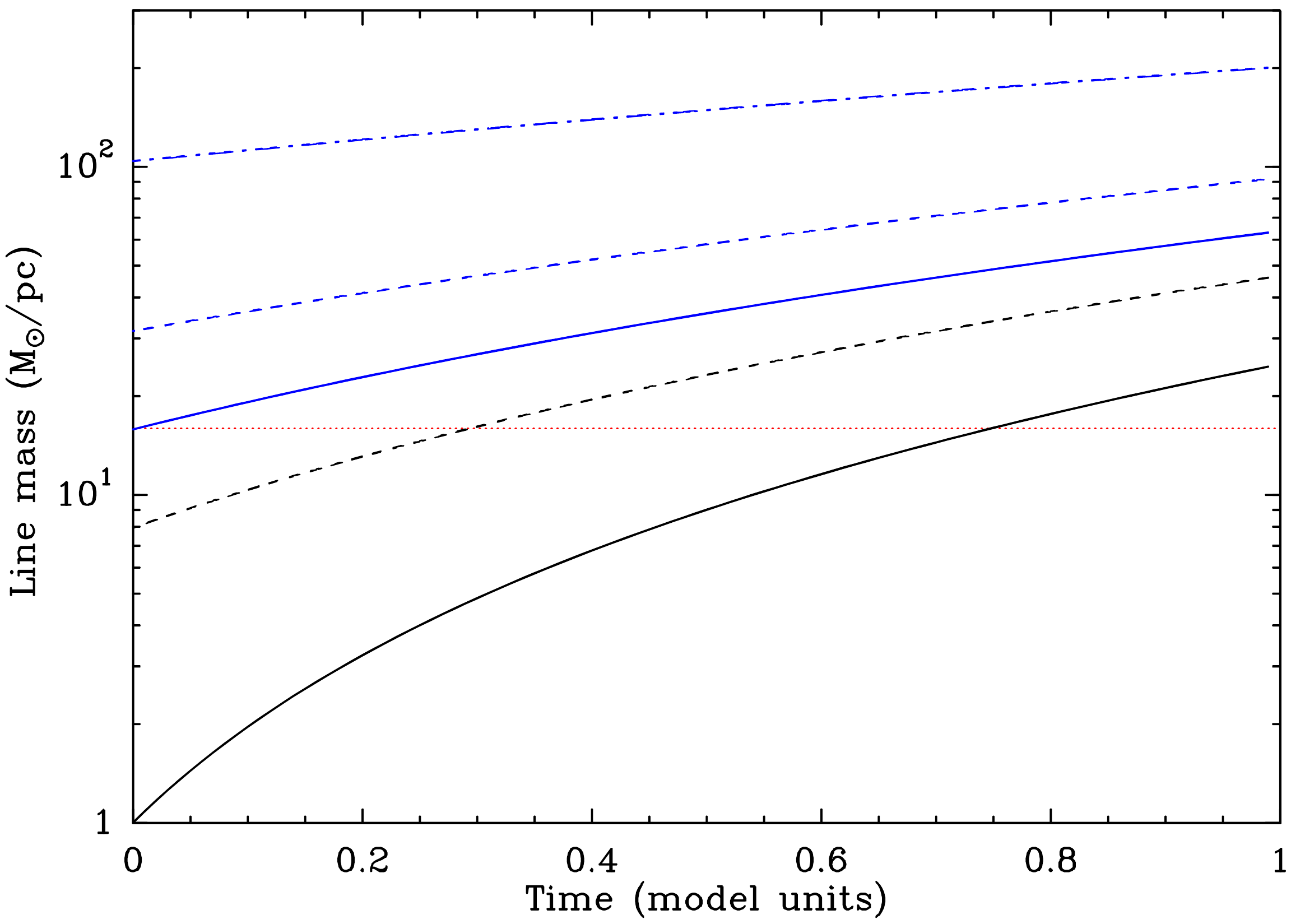}}}            
\caption{Evolution of the mass per unit length of filaments according to the toy gravitational-accretion model 
described in the text, for five values of the initial line mass at $t=0$: 
$M_{\rm line}(0) = 1\, M_\odot$/pc (black solid curve), $M_{\rm line}(0) \sim 8\, M_\odot$/pc (half critical, black dashed curve), 
$M_{\rm line}(0) \sim 16\, M_\odot$/pc (critical, blue solid curve), $M_{\rm line}(0) \sim 32\, M_\odot$/pc (twice critical, blue dashed curve), 
$M_{\rm line}(0)  \sim 100\, M_\odot$/pc (highly supercritical, blue dash-dotted curve). 
The red dotted horizontal line marks the critical line mass $ \sim 16\, M_\odot$/pc. 
Time is normalized in such a way that a critical filament with $M_{\rm line} = M_{\rm line, crit}$ at $\tilde{t} = 0$ has 
$M_{\rm line} = 4 \times M_{\rm line, crit}$ at $\tilde{t} = 1$ (cf. solid blue curve), corresponding to $\sim\,$1--2~Myr.
}
\label{model_lmass_evol}
\end{figure}

In absolute terms, the characteristic accretion timescale is
on the order of 1--2~Myr for a Taurus-like filament with $ M_{\rm line} \sim 50\, M_\odot $/pc$\ \sim 3\, M_{\rm line, crit}$ \citep[][]{Palmeirim+2013, Shimajiri+2019}.
Figure~\ref{model_lmass_evol} shows the time evolution $ M_{\rm line}(\tilde{t}) $ predicted by this simple accretion model 
for five values of the initial line mass $M_{\rm line}(0)$.

In the context of this model, we may derive the time evolution of the FLMF 
following an approach similar to that employed by \citet{Zinnecker1982} 
in his toy model of the IMF based on Bondi-Hoyle accretion (for which $\dot{M}_\star \propto M_\star ^2 $). 
Mass conservation implies that the cumulative distribution 
of line masses at time  $\tilde{t} $, $N_{\tilde{t}}\, \left[  > M_{\rm line} \right] $ is related to 
the initial distribution of line masses by:

\begin{equation}
N_{\tilde{t}}\, \left[ > M_{\rm line} \right] 
= N_0  \left[ \, > M_{0,\tilde{t}}\,(M_{\rm line}) \right] \equiv  \int_{M_{0,\tilde{t}}\,(M_{\rm line})}^{+\infty} \frac{ {\rm d} N_0 }{{\rm d} M_{\rm line}^0 }  \, {\rm d} M_{\rm line}^0   \,,
\end{equation}

\noindent 
where $M_{0,\tilde{t}}\, (M_{\rm line})  =  \left(M_{\rm line}^{1/2}- M_{\rm line, crit}^{1/2}\, \tilde{t}\,  \right) ^2$ represents the initial line mass of a filament with line mass $M_{\rm line}$ at time $\tilde{t}$. 
The differential FLMF at time $\tilde{t} $ can then be obtained by taking the derivative of Eq.~(C.5) with respect to $M_{\rm line} $: 

\begin{equation}
\frac{ {\rm d} N_{\tilde{t}} }{{\rm d} M_{\rm line}} \equiv -\frac{ {\rm d} N_{\tilde{t}}\, \left[ > M_{\rm line} \right] }{{\rm d} M_{\rm line}} 
= \frac{ {\rm d} N_0 }{{\rm d} M_{\rm line}^0}\, \left[ M_{0,\tilde{t}}\, (M_{\rm line})  \right]  \times \frac{ {\rm d} M_{0,\tilde{t}} }{{\rm d} M_{\rm line}}    \,,
\end{equation} 

\noindent
which leads to:

\begin{equation}
\frac{ {\rm d} N_{\tilde{t}} }{{\rm d} M_{\rm line}}\, (M_{\rm line}) = \frac{ {\rm d} N_0 }{{\rm d} M_{\rm line}^0}\, \left[ M_{0,\tilde{t}}\, (M_{\rm line})  \right]  \times \left(1-\frac{M_{\rm line, crit}^{1/2}}{M_{\rm line}^{1/2}}\,\tilde{t} \right)  \,.
\end{equation} 

\noindent
The latter can also be written as:

\begin{equation}
\frac{ {\rm d} N_{\tilde{t}} }{{\rm d} M_{\rm line}}\, (M_{\rm line}) = \frac{ {\rm d} N_0 }{{\rm d} M_{\rm line}^0}\, \left[ M_{0,\tilde{t}}\, (M_{\rm line})  \right]  
\times \left( \frac{M_{\rm line}}{M_{0,\tilde{t}}} \right)^{1/2} \,.
\end{equation} 

\begin{figure}[!ht]
\centerline{\resizebox{0.95\hsize}{!}{\includegraphics[angle=0]{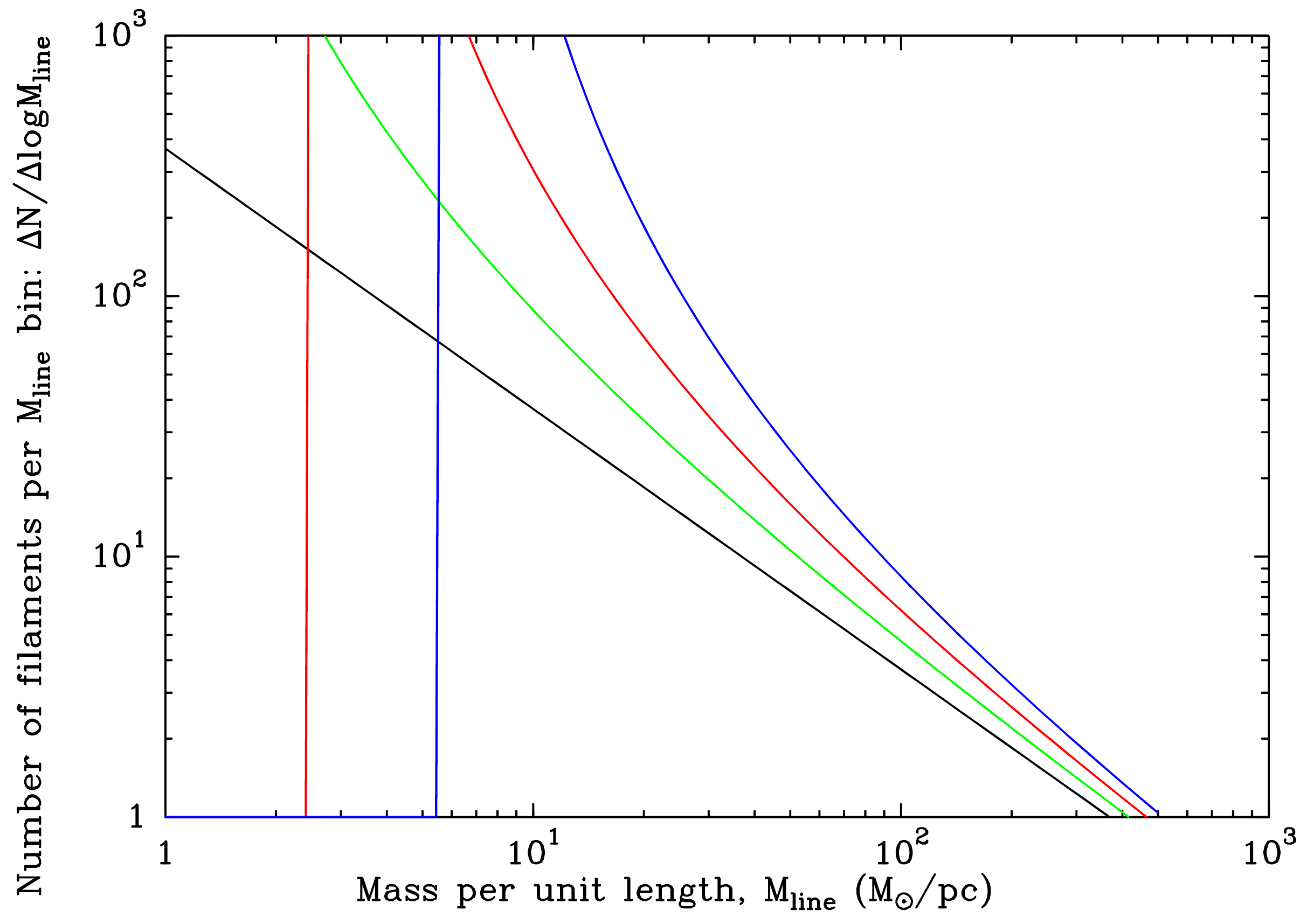}}}            
\caption{Evolution of the FLMF in the context of the proposed toy accretion model. 
The black solid line shows the initial power-law FLMF  (d$N$/dlog$M_{\rm line} \propto M_{\rm line}^{-1}$)
determined by 
interstellar turbulence. 
The green, red, and blue solid curves show the model FLMF at three time steps, $\tilde{t}=0.2$, $\tilde{t}=0.4$, 
$\tilde{t}=0.6$ after the accretion process is ``switched on'' at $\tilde{t}=0$, where $\Delta \tilde{t}=0.4$ roughly 
corresponds to the time it takes for a critical filament to double its mass per unit length ($\sim \,$0.5--1~Myr). 
The median logarithmic slope of the model FLMF for $16 < M_{\rm line} < 500\, M_\odot $/pc is $-1$, $-1.14$, $-1.30$, and $-1.50$
at $\tilde{t}=0$, $\tilde{t}=0.2$, $\tilde{t}=0.4$, and $\tilde{t}=0.6$, respectively.
The vertical red and blue lines correspond to the line mass $M_{\rm line, crit}\, \tilde{t}^2$ accreted 
by filaments with $M_{\rm line}(0)  \approx 0$ at $\tilde{t}=0.4$ and $\tilde{t}=0.6$, respectively. 
}
\label{model_flmf_nodecay}
\end{figure}

Starting from an initial power-law FLMF d$N$/dlog$M_{\rm line} \propto M_{\rm line}^{-1}$ 
determined by 
interstellar turbulence (cf. Sect.~\ref{sec:concl}), 
the resulting FLMF is shown at three time steps, $\tilde{t}=0.2$, $\tilde{t}=0.4$, $\tilde{t}=0.6$, and compared 
to the initial power law at $\tilde{t}=0$ in Fig.~\ref{model_flmf_nodecay}. 
It can be seen that the accretion process steepens the model FLMF with time, 
making it more consistent with the observed FLMF of Fig.~\ref{fmf}a in the supercritical regime 
than the initial power-law FLMF. 
In particular, the median logarithmic slope of the model FLMF 
in the range of line masses $16 < M_{\rm line} < 500\, M_\odot $/pc is between $-1.5$ and $-1.3$, 
i.e., Salpeter-like at $\tilde{t}\sim \,$0.2--0.4 (i.e., less than 1~Myr  after the onset of accretion), 
in good agreement with the observed FLMF which has a logarithmic slope of $-1.4 \pm 0.1$ (Fig.~\ref{fmf}a). 

\begin{figure*}[!htp]
\centerline{\resizebox{0.48\hsize}{!}{\includegraphics[angle=0]{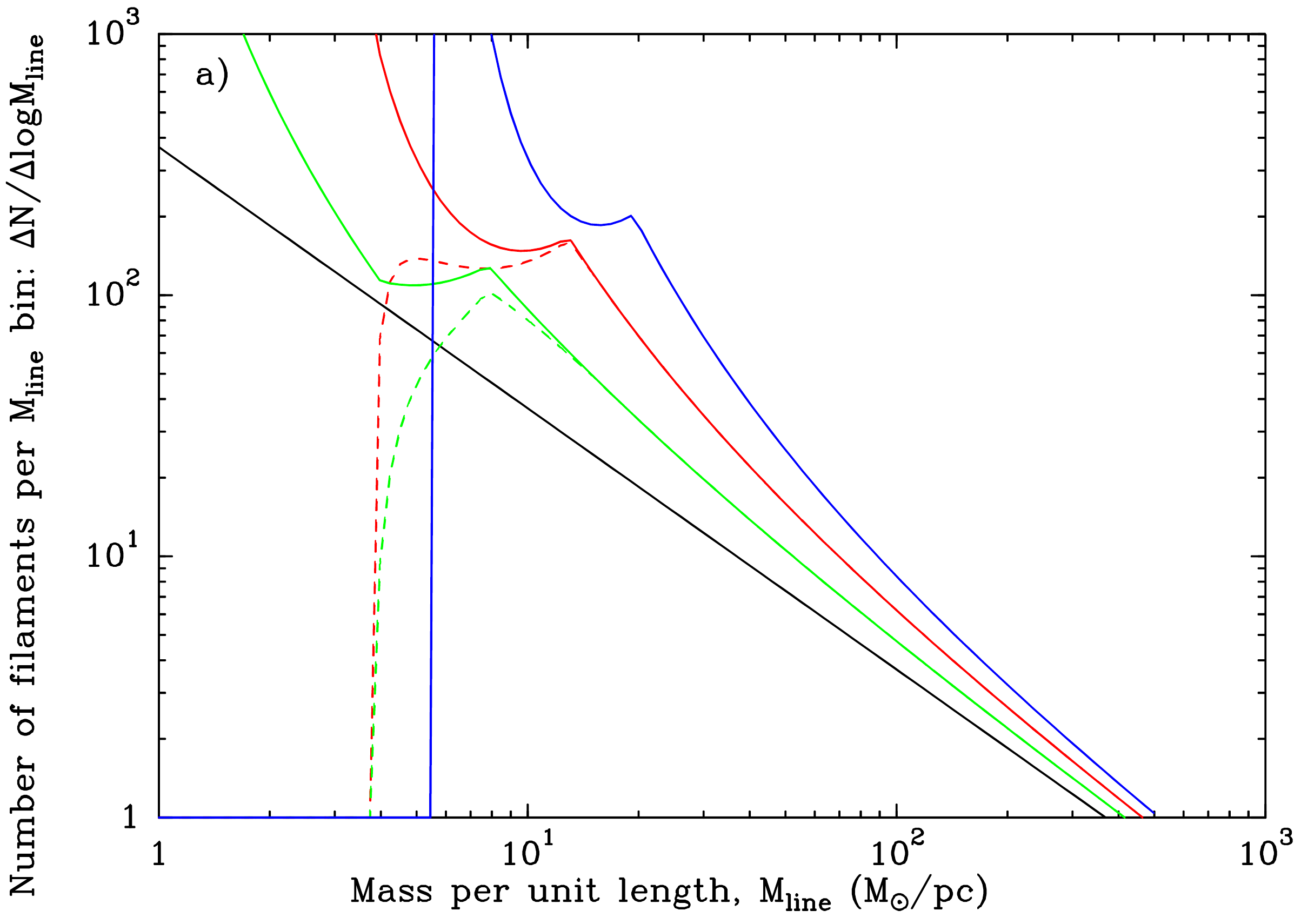}}
            \hspace*{0.3cm}
            \resizebox{0.48\hsize}{!}{\includegraphics[angle=0]{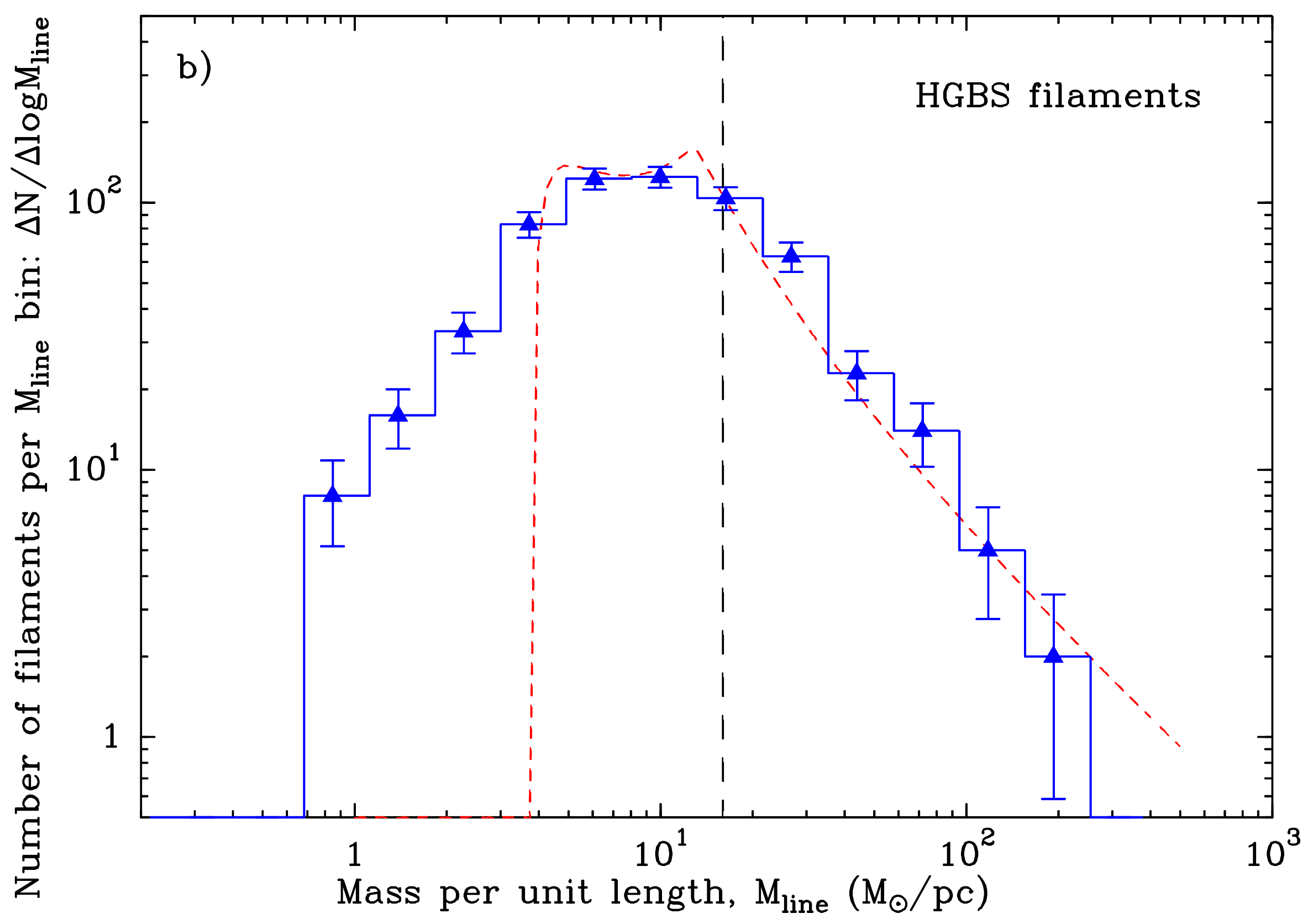}}}            
\caption{{\bf a)} Evolution of the FLMF according to the improved version of our model, 
where filaments accrete in the same way as in Fig.~\ref{model_flmf_nodecay},  
but subcritical filaments with $ M_{\rm line} < 4\, M_\odot $/pc decay on a timescale $\tilde{\tau}_{\rm decay} = 0.15 $ at the same time. 
The black solid line shows the initial power-law FLMF  (d$N$/dlog$M_{\rm line} \propto M_{\rm line}^{-1}$)
determined by 
interstellar turbulence. 
The green, red, and blue solid curves show the model FLMF at three time steps, $\tilde{t}=0.2$, $\tilde{t}=0.4$, 
$\tilde{t}=0.6$ after the accretion process is ``switched on'' at  $\tilde{t}=0$, where $\tilde{t}=0.4$ roughly corresponds to $\sim \,$0.5--1~Myr. 
The vertical red and blue lines are the same as in Fig.~\ref{model_flmf_nodecay}. 
The green, red, and blue dashed curves show the same model FLMF taking 
estimated incompleteness effects into account in the subcritical line mass regime. 
{\bf b)}  Comparison of the model FLMF including the incompleteness effect at $\tilde{t}=0.4$ (dashed red curve) 
with the observed FLMF from Fig.~\ref{fmf}a  (blue histogram). 
}
\label{model_flmf}
\end{figure*}

The model FLMF nevertheless quickly diverges near $ M_{\rm line} = M_{\rm line, crit}\, \tilde{t}^2$, 
due to an accumulation of filaments with very low initial masses per unit length, i.e., $M_{\rm line}(0)  \approx 0$, 
whose $ M_{\rm line}(\tilde{t})$ is entirely built up by gravitational accretion. 
This is not very physical since filaments that are highly subcritical initially ($ M_{\rm line}(0)  <<  M_{\rm line, crit}/2 $) 
are not self-gravitating and are unlikely to gravitationally accrete mass from the ambient cloud. 
Instead, these filaments may disperse on a sound crossing time unless they are pressure-confined. 
In Fig.~\ref{model_flmf}, we present an improved version of the same accretion model where 
the number of subcritical filaments with $ M_{\rm line} < 4\, M_\odot $/pc decay 
on a characteristic timescale $\tilde{\tau}_{\rm decay} = 0.15 $, at the same time 
as the filaments accrete mass on the timescale $\tilde{\tau}_{\rm acc} $ given by Eq.~(C.4). 
It can be seen that this modified model provides a better match 
to the observed FLMF (cf. Fig.~\ref{model_flmf}b), especially when incompleteness effects 
are taken into account  in the subcritical line mass regime (cf. dashed curves in Fig.~\ref{model_flmf}a).

\end{appendix}

\end{document}